\documentclass[namedreferences]{SolarPhysics}
\usepackage[optionalrh,solaenum,natbib]{spr-sola-addons} % For Solar Physics 
\usepackage{graphicx}                    % For eps figures, newer & more powerfull
\usepackage{amssymb}                    % useful mathematical symbols
\usepackage{color}                       % For color text: \color command
\usepackage{url}                         % For breaking URLs easily trough lines
\usepackage{sistyle}			% Additional support for SI symbols etc.

\begin{document}

\begin{article}

\begin{opening}

\title{Is Cycle 24 the Beginning of a Dalton-Like Minimum?}

%%%%%%%%%%%%%%%%%%%%%%%%%%%%%%%%%%%%%%%%%%%%%%%%%%%
%% Authors Names
%
\author{M.~\surname{Lindholm Nielsen}$^{1}$\sep
        H.~\surname{Kjeldsen}$^{1}$\sep
%        I.~\surname{}$^{2}$      
       }

%%%%%%%%%%%%%%%%%%%%%%%%%%%%%%%%%%%%%%%%%%%%%%%%%%%
%% Runningheads
%
%\runningauthor{}
%\runningtitle{}

%%%%%%%%%%%%%%%%%%%%%%%%%%%%%%%%%%%%%%%%%%%%%%%%%%%
%% Affilations 
%
  \institute{$^{1}$ Aarhus University
                     email: \url{mln@phys.au.dk} \\ 
             $^{2}$ Aarhus University
                     email: \url{hans@phys.au.dk} \\
             }

%%%%%%%%%%%%%%%%%%%%%%%%%%%%%%%%%%%%%%%%%%%%%%%%%%%
%% Abstract 
\begin{abstract}
The unexpected development of cycle 24 emphasizes the need for a better way to model future solar activity. 
In this article, we analyze the accumulation of spotless days during individual cycles from 1798-2010. The analysis shows that spotless days do not disappear abruptly in the transition towards an active sun. 
A comparison with past cycles indicates that the ongoing accumulation of spotless days is comparable to that of cycle 5 near the Dalton minimum and to that of cycles 12, 14 and 15.
%A comparison with past cycles indicates that the accumulation of spotless days during the ongoing cycle is comparable to that of cycle 5 near the Dalton minimum and to that of cycles 12,14 and 15. 
It also suggests that the ongoing cycle has as much as $20\pm8$ spotless days left, from July 2010, before it reaches the next solar maximum. The last spotless day is predicted to be in December 2012, with an uncertainty of 11 months. This trend may serve as input to the solar dynamo theories.
\end{abstract}

%%%%%%%%%%%%%%%%%%%%%%%%%%%%%%%%%%%%%%%%%%%%%%%%%%%
%% Keywords
%

\end{opening}
%-------------------------------------------------

%%%%%%%%%%%%%%%%%%%%%%%%%%%%%%%%%%%%%%%%%%%%%%%%%%%
%% Sections
%
\section{Introduction}%\label{s:?} 
The sun is a variable star, characterized by the quasi-periodic Schwabe cycle. The variation is, among other things, observed by the changing number of sunspots, changes in the total solar irradiation and by the varying charged particle flux  from the sun \citep{Activity}. Understanding and predicting this activity is of prime importance to the protection of satellites, astronauts, as well as ground based electrical installations \citep{MagneticStorm}.

%It has however proved difficult to predict the level of future solar activity, both from a theoretical point of view, as well as from empirical observations. 
The activity is driven by the solar dynamo, for which we have no complete understanding \citep{Dynamo}.  Nevertheless, there have been several attempts to extrapolate future solar activity, based on models and observations \citep{Models1}. This also includes the use of spotless days as a predictor for future solar activity \citep{Wilson,NASA1,NASA2,NASA3,NASA4}. The prediction accuracy varies from model to model, as is clearly demonstrated by the recent extended minimum \citep{Models2}. However, so far no one has used the accumulative number of spotless days to predict future solar activity.

Here we use the Wolf/Zurich/International (WZI) and Hoyt \& Schatten (HS) sunspot records as a basis for characterizing intervals with low solar activity. The cycle minima are found using a Gaussian filter. Next the accumulation of spotless days during individual cycles is analyzed. The analysis includes a condition that ensures a stable starting point for the accumulative number of spotless days. Based on this condition we compare and discuss the ongoing cycle in relation to earlier similar cycles.

\section{The Data}
The sunspot data are available via the National Geophysical Data Center \footnote{The sunspot data was downloaded on July 15, 2010~at \url{http://www.ngdc.noaa.gov/stp/solar/ssndata.html}}. Two different tables of sunspot data have been used. 
The first is a collection of WZI sunspot records observed daily since 1818. This collection is quantified by the \emph{relative sunspot number} which is an index that describes the activity of the sun's visible disk and is defined as
\begin{equation}
R_\textrm{w} = K(10g + f)	
\label{R_w}
\end{equation}
\noindent where $K$ is a reduction coefficient that depends on the observational method, $g$ is the number of sunspot groups and $f$ is the total number of spots on the visible disk.

% Figure 

\begin{figure}
\centering
\includegraphics[width=\columnwidth]{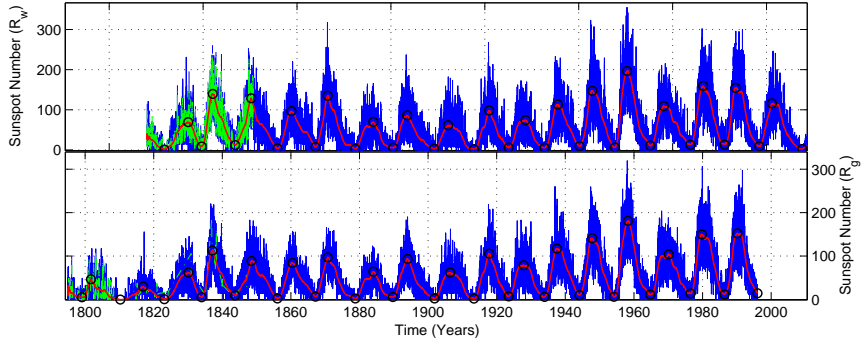}
 \caption{WZI time series (top) and HS time series (bottom). Here blue and green curves show the original and interpolated data, respectively. The red curves represent the data smoothed with a Gaussian of full width at half maximum (FWHM) of \SI{506}{days}. The black circles denote extrema points near sunspot minima and maxima.}
 \label{fig:sunspots}
 \end{figure}

The second dataset is a collection of daily \emph{sunspot group numbers} spanning from 1610 to 1995. These data come from 463 independent observers and give a more complete picture of solar activity on longer timescales \citep{Hoyt1,Hoyt2}.
From their analysis of historical records Hoyt and Schatten were able to construct the sunspot group number and make it compatible with the relative sunspot number using a normalization procedure. With this normalization the index is defined as
\begin{equation}
R_\textrm{g} = \frac{12.08}{N} \sum_{i = 1}^{N}k_ig_i 
\label{R_g}
\end{equation}

\noindent where 12.08 is the normalization number which makes the mean $R_\textrm{g}$'s identical with the mean $R_\textrm{w}$'s for 1874 to 1976 when the Royal Greenwich Observatory (RGO) actively made observations of sunspot groups. $N$ is the number of observers, $k_i$ is a correction factor, and $g_i$ is the number of sunspot groups as reported by observer $i$. The correction factor $k_i$ is defined as 1.000 for the primary observer RGO. For all other observers the correction factor is constructed by dividing the total number of sunspot groups as reported by the comparison observer and by RGO. The details surrounding the creation of the correction factor on data before 1874 are somewhat different and is discussed at length by \citet{Hoyt1,Hoyt2}.

\section{Analysis}
\begin{table}
\caption{This table illustrates low active intervals based on the zero point condition described in the text. The first column highlight the respective time series, the second is the interval of interest, the third display the related cycle number and the forth column show the corresponding minimum which was found using the Gaussian filter. TSD is the total number of spotless days in the interval and IP is the corresponding interpolation. The last column show the estimated accumulated error in the total number of spotless days.}
\label{tbl:timeseries}
\begin{tabular}{lllllll}
\hline
\textbf{TS}&\textbf{Interval}&\textbf{Cycle}&\textbf{Minimum}&\textbf{TSD}&\textbf{IP ($\%$)}&\textbf{Error ($\%$)}\\
\hline
\textbf{WZI}&2007.12.19 - 2010.6.16&C24&2009.2.8&571&0&0\\
&1996.7.4 - 1998.1.9&C23&1996.7.21&152&0&0\\
&1986.6.2 - 1987.7.14&C22&1986.6.7&125&0&0\\
&1976.5.25 - 1977.7.18&C21&1976.6.16&89&0&0\\
&1964.6.5 - 1966.8.10&C20&1964.10.8&168&0&0\\
&1953.10.18 - 1955.10.18&C19&1954.6.17&344&0&0\\
&1943.12.26 - 1945.9.15&C18&1944.2.19&181&0&0\\
&1933.3.13 - 1935.7.29&C17&1933.12.11&390&0&0\\
&1923.1.5 - 1926.7.18&C16&1923.6.21&347&0&0\\
&1912.9.22 - 1916.10.2&C15&1913.5.14&548&0&0\\
&1900.10.31 - 1905.7.28&C14&1901.9.20&639&0&0\\
&1889.8.20 - 1891.12.17&C13&1889.10.27&280&0&0\\
&1877.9.22 - 1883.9.25&C12&1878.10.12&600&0&0\\
&1866.11.4 - 1869.7.14&C11&1867.2.22&303&0&0\\
&1855.11.3 - 1858.4.5&C10&1856.1.27&380&0&0\\
&1843.6.6 - 1847.7.4&C9&1843.9.26&259&25.44&7.797\\
&1833.12.17 - 1835.6.19&C8&1833.12.19&198&32.18&29.86\\
\textbf{HS}&1986.6.2 - 1987.7.14&C22&1986.6.5&125&0&0\\
&1976.5.25 - 1977.7.18&C21&1976.7.3&91&0&0\\
&1964.6.5 - 1966.8.10&C20&1964.10.8&205&0&0\\
&1953.10.18 - 1955.10.18&C19&1954.3.3&349&0&0\\
&1943.12.26 - 1945.9.15&C18&1944.2.12&179&0&0\\
&1933.3.13 - 1935.7.29&C17&1933.12.11&396&0&0\\
&1923.1.5 - 1926.7.18&C16&1923.6.21&339&0&0\\
&1912.9.21 - 1916.10.2&C15&1913.5.14&553&0&0\\
&1900.10.31 - 1903.9.22&C14&1901.11.21&666&0&0\\
&1889.8.21 - 1891.3.27&C13&1889.9.23&269&0&0\\
&1877.9.22 - 1883.5.27&C12&1878.10.12&595&0&0\\
&1866.12.2 - 1868.7.13&C11&1867.2.22&262&0&0\\
&1855.11.3 - 1858.1.17&C10&1856.1.28&425&0&0\\
&1843.6.6 - 1847.7.30&C9&1843.9.23&270&2.704&0.7554\\
&1833.2.26 - 1835.6.19&C8&1833.12.20&437&0.8294&0.4458\\
&1822.4.12 - 1830.1.24&C7&1823.2.5&982&1.265&1.283\\
&1808.6.10 - 1816.8.27&C6&1810.4.10&1905&3.732&0.8849\\
&1798.4.1 - 1801.6.10&C5&1799.1.25&708&23.84&23.85\\
\hline
\end{tabular}
\end{table}
It is important to understand the conditions and limitations of each time series. Regarding the WZI time series, \citet{Hoyt1,Hoyt2} point out that Wolf relied upon correspondents to analyze and send the results to him. The quality of these interpretations was sometimes poor because of an unclear definition of a group and a sunspot. This is clear from a reexamination of original observations from 1819-1833 which shows that some individual sunspots were counted as sunspot groups, thus leading to erroneous higher values of the relative sunspot number \citep{Hoyt1995}. In addition the methodology changed in 1882 where smaller spots were counted and the reduction coefficient $K$ changed from $K=1$ to $K=0.6$ \citep{Waldmeier}. Furthermore, before 1982 the index was calculated using a single observer which should give rise to a more homogenous time series. As a drawback, all other observations were neglected making observational uncertainties difficult to estimate. Since 1982 the index has been calculated using 20 independent observers \citep{Activity}.
There is no missing observational data after 1849, but in the years between 1818 - 1849 we have used linear interpolation to make up for missing data.

With respect to the HS time series, \citet{Hoyt1,Hoyt2} used linear interpolation to fill gaps of up to four days for an active Sun and six days for a quiet Sun. There is no missing observational data after 1848, but in the years between 1798 - 1848 we have used linear interpolation to make up for missing data.
Moreover, because of large gaps in the time series, the analysis is restricted to the period after 1798.

To locate the time index near sunspot minima and maxima, we use a Gaussian filter, with a full width at half maximum (FWHM) of \SI{506}{days}.
This smooths the time series enough to locate extrema points near sunspot minima and maxima.
We then isolate individual cycles based on the located maxima points. The two time series, located extrema points and associated cycles are shown in Figure \ref{fig:sunspots}. In addition, the intervals of interest, related cycle numbers, associated minima and amount of interpolation performed are outlined in Table \ref{tbl:timeseries}.

\subsection{Determining the Zero Points for Low Activity}
\begin{figure}
\centering
\includegraphics[width=\columnwidth]{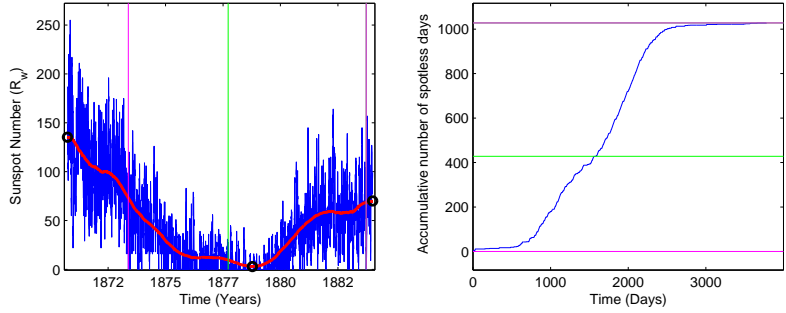}
 \caption{Segment of the WZI time series covering cycle 12 (left) and the accumulated number of spotless days during cycle 12 (right). The vertical magenta lines mark the first and last spotless day. The corresponding accumulation of spotless days, from the first to the last spotless day, is marked by magenta horizontal lines in the right panel. The vertical green line marks the zero point spotless day which has $\Delta T \le 14$ days to the previous spotless day. The corresponding cutoff is shown by the horizontal green line in the right panel.}
 \label{fig:condition}
 \end{figure}
It is of interest to isolate the onset of low solar activity in a given cycle. To do this we use the spotless days as a proxy for low activity.
The time between spotless days varies from months to weeks leading to spotless streaks near solar minima. 
To find the onset of spotless streaks in a given cycle, we isolate all the spotless days between the first relevant sunspot maxima and the following minima.
%To find the onset of spotless streaks we first isolate all the spotless days between the relevant sunspot maxima and minima. 
Next, we calculate the time $\Delta T$ between the neighboring spotless days. Finally, we find the last spotless day in this interval, which has $\Delta T \leq \SI{14}{~days}$ to the previous spotless day. This isolated day is then used as the zero point for the cumulative summation of spotless days during the relevant cycle. The condition $\Delta T \leq \SI{14}{~days}$ was found such that the zero point lies before the respective minima and is close to the onset of the first spotless streak. The procedure is illustrated for WZI cycle 12 (C12) in Figure \ref{fig:condition}, where the cutoff in the right panel, marks the respective zero point of interest. Repeating the procedure for all cycles leads to the final result shown in Figure \ref{fig:analysis}. 

%The procedure is illustrated in Figure \ref{fig:condition} and the final result is shown in Figure \ref{fig:analysis}. 

\subsection{Correspondence between Spotless Days from the HS and WZI Time Series}
\begin{figure}[ht]
\centering
\includegraphics[width=\columnwidth]{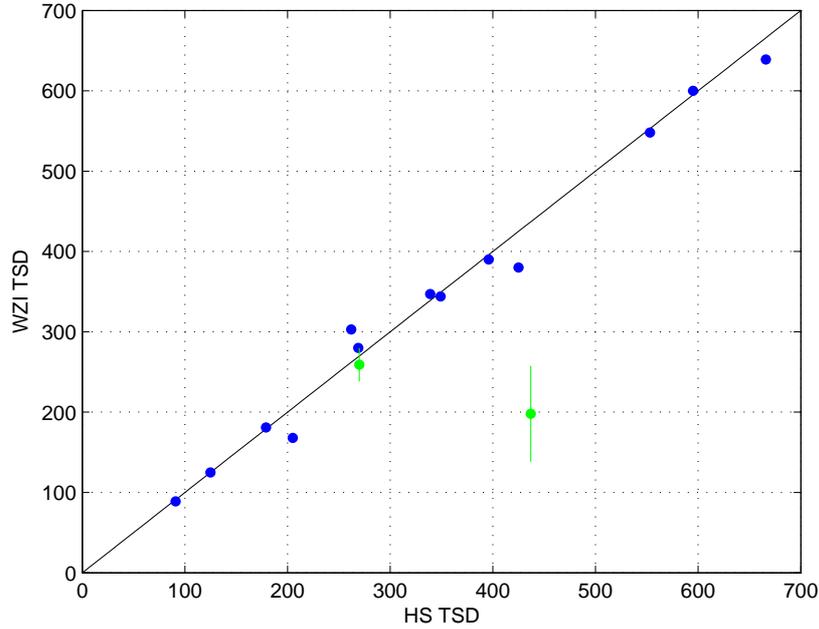}
 \caption{The total number of spotless days (TSD) during the intervals specified in Table \ref{tbl:timeseries}, from the WZI time series plotted against similar results from the HS time series. Here blue show the TSD for cycles C10-C22 with no interpolation and green show the TSD for cycles C8-C9 containing interpolation.
The scaling is roughly similar in the two records except for the outlier of cycle 8.}
 \label{fig:qq}
 \end{figure}
\begin{figure}
 \centerline{\includegraphics[width=\columnwidth]{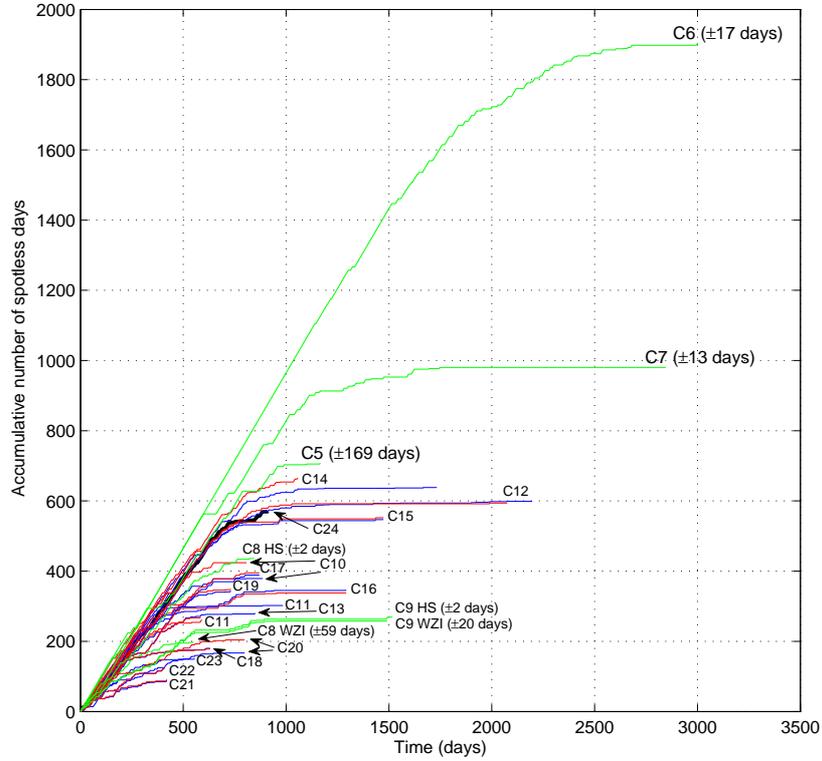}}
 \caption{The accumulative number of spotless days, counted from the first spotless day which was found using the zero point condition described in the text. Here red and blue show the HS and WZI time series, respectively. The interpolated intervals from both time series is shown in green with the estimated accumulated error displayed in the adjacent parenthesis.  %Notice how the ongoing cycle compares to C5 near the Dalton minimum and to cycles 12,14 and 15.
}
 \label{fig:analysis}
 \end{figure}

It is of interest to compare the ongoing accumulation of spotless days for cycles near the Dalton minimum. 
However, the WZI time series only extends back to 1818 and there are gaps of missing data between 1818-1849. Moreover, as previously mentioned, the reliability of the WZI time series before 1834 is also in question \citep{Hoyt1995}. Given that the HS time series ended in 1995 a direct comparison, to the ongoing accumulation, relies on the correspondence between spotless days from the HS and WZI time series.
The correspondence is best illustrated by plotting the total number of spotless days (TSD), in a given cycle, from the WZI time series against similar results from the HS time series. 
%The correspondence is best illustrated in a q-q plot where quantiles of the WZI data set is plotted against quantiles of the HS data set. The quantiles are in this case the total number of spotless days in a given cycle.
The relevant data are outlined in Table \ref{tbl:timeseries} and illustrated in Figure \ref{fig:qq}. This figure shows that the scaling of TSD is roughly similar in the two records, except for cycle 8 which may be due to the uncertain reliability of the WZI time series before 1834. 
Moreover, the change in methodology in the WZI time series, does not seem to affect the number of spotless days after 1882. Thus, the accumulated number of spotless days from the two time series are combined in Figure \ref{fig:analysis}.

% Figure 
 
 \begin{figure} 
 \centerline{\includegraphics[width=\columnwidth]{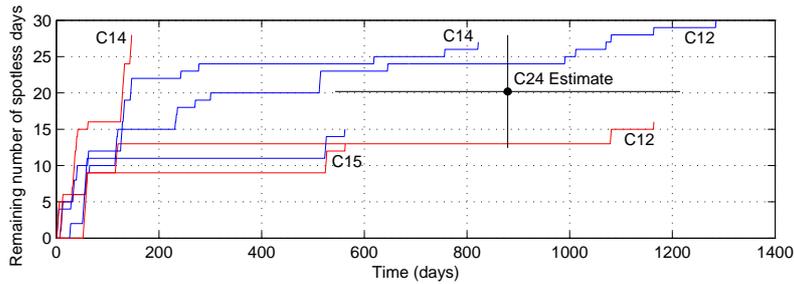}}
 \caption{The remaining number of spotless days, counted from the last spotless day of C24 in Figure \ref{fig:analysis}. Here red and blue show cycles 12, 14 and 15 from the HS and WZI time series, respectively. The  estimate on the remaining number of spotless days in C24 is marked in black. Note that on the last spotless day of WZI C14 (28th of July 1905) $R_\textrm{w} = 0$ while $R_\textrm{g} = 1\pm4$. Thus, because of this large uncertainty HS C14 is neglected in the estimate of C24.}
 \label{fig:estimate}
 \end{figure}
 
\subsection{Estimating the Accumulated Error}
It is of interest to estimate the accumulated error in the TSD which arise from the linear interpolation performed in some intervals.
To estimate the error, we temporarily remove similar amounts of data in later non-interpolated cycles. Next, we employ linear interpolation over this temporary data and calculate the TSD using the zero point condition described previously. The result is then compared to the non-interpolated TSD in the respective interval. Finally, the average deviation is calculated and used to quantify the accumulated error in the real interpolated interval. The results are again outlined in Table \ref{tbl:timeseries} and are also shown in the parenthesis in Figure \ref{fig:analysis}.  

\section{Discussion}
The analysis shows that spotless days do not disappear abruptly in the transition toward an active sun, but rather there is a gradual decline in the number of spotless  days. This is also the case for highly active cycles, although the rate of decline seems to differ somewhat. Figure \ref{fig:analysis} shows the ongoing accumulation of spotless days compared to previous cycles. 
It is noteworthy that the ongoing accumulation of C24 is comparable to C5 near the Dalton minimum and to cycle 12,14 and 15 near the modern minimum \citep{Activity}.
The cycles following C5 and C14 suggest that C25 will be even less active than C24 and this trend may serve as input to the solar dynamo theories.
 
%In comparing the different cycles, one should note that the level of activity can be lower than what is indicated by a spotless day. This is illustrated by the \SI{10.7}{cm} radio observations of the sun which differs between low active intervals. This means that the sunspot number can only be used as a relative indicator of low solar activity. 

To estimate the remaining number of spotless days in C24, we shift the zero point of Figure \ref{fig:analysis} by 911 days, which corresponds to the present accumulation of C24. The result is shown in Figure \ref{fig:estimate} which illustrates the remaining number of spotless days for cycles C12, C14 and C15.
There is an obvious difference in the last recorded spotless day of C14. From the original data, we note that on the last spotless day of WZI C14 (28th of July 1905) $R_\textrm{w} = 0$ while $R_\textrm{g} = 1\pm4$. Thus, because of this large uncertainty, we neglect HS C14 in the following calculation. 
Based on the data related to Figure \ref{fig:estimate} we find that $20\pm8$ spotless days remain, from July 2010, before the next solar maximum. The last spotless day is predicted to occur in December 2012 with an uncertainty of 11 months.
The large uncertainty is related to the difference between C12 and C15.
Evidently C12 is special in the sense that the last spotless day appears more than $600$ days later relative to C15. However, C12 is retained in our estimate because of its close similarity to C24 shown in Figure \ref{fig:analysis}.

Previously, \citet{NASA1} have shown that the time from the first to the last spotless day is strongly correlated ($r = 0.949$ with a standard error of estimate (se) equal to 9.9 months) with the time from first spotless day and sunspot minimum occurrence. For C24 the first spotless day occurred in January 2004, while a 12-month moving average yields a sunspot minimum in December 2008. Thus, the time between the first spotless day and minimum occurrence is $\approx 60$ months. Therefore, the expected interval of spotless days for C24 is $90.7\pm9.9$ months. The time from first spotless day occurrence to July 2010 is 78 months. Hence, the last spotless day is expected to occur within the next 22 months from July 2010, or before May 2012. 
It should be noted that C12 is shown to be an outlier of $\approx 20$ months in the analysis of \citet{NASA1}. Thus, there is overlap between the two estimates.

\section*{Acknowledgements}
We would like to thank J\o rgen Christensen-Dalsgaard for extensive discussions in relation to the present project.

% \section{}%\label{s:?} 

%% Figure 
%
% \begin{figure} 
% \centerline{\includegraphics[width=0.5\textwidth,clip=]{<fig.eps>}}
% \caption{}%\label{fig:?}
% \end{figure}

%% Table
%
% \begin{table}
% \caption{}%\label{tbl:?}
% \begin{tabular}{}     
% \hline
% \multicolumn{2}{c}{<>}
% <data>
% \hline
% \end{tabular}
% \end{table}

%%%%%%%%%%%%%%%%%%%%%%%%%%%%%%%%%%%%%%%%%%%%%%%%%%%%%%%%%%%%%%%%%%%%%%%%%%%
%% Appendix
%
% \appendix   

%%%%%%%%%%%%%%%%%%%%%%%%%%%%%%%%%%%%%%%%%%%%%%%%%%%%%%%%%%%%%%%%%%%%%%%%%%%
%% Acknowledgements
%
% \begin{acks}
%
% \end{acks}

%%% %%%%%%%%%%%%%%%%%%%%%%%%%%%%%%%%%%%%%%%%%%%%%%%%%%%%%%%%%%%
%% Bibliography
%
% Using BibTeX
%
% \bibliographystyle{sola-support/spr-mp-sola}
\bibliographystyle{sola-support/spr-mp-sola-cnd} %% Alternative style: no title, no concluding page
\bibliography{oldbib}  

\begin{thebibliography}{14}
% BibTex style file: spr-mp-sola-cnd.bst, 2010-05-13
\ifx \bisbn   \undefined \def \bisbn  #1{ISBN #1}\fi
\ifx \binits  \undefined \def \binits#1{#1}\fi
\ifx \bauthor  \undefined \def \bauthor#1{#1}\fi
\ifx \batitle  \undefined \def \batitle#1{#1}\fi
\ifx \bjtitle  \undefined \def \bjtitle#1{\textit{#1}}\fi
\ifx \bvolume  \undefined \def \bvolume#1{\textbf{#1}}\fi
\ifx \byear  \undefined \def \byear#1{#1}\fi
\ifx \bissue  \undefined \def \bissue#1{#1}\fi
\ifx \bfpage  \undefined \def \bfpage#1{#1}\fi
\ifx \blpage  \undefined \def \blpage #1{#1}\fi
\ifx \burl  \undefined \def \burl#1{\textsf{#1}}\fi
\ifx \href  \undefined \def \href#1#2{\textsf{#2}}\fi
\ifx \doiurl  \undefined \def
  \doiurl#1{\href{http://dx.doi.org/#1}{\textsf{#1}}}\fi
\ifx \betal  \undefined \def \betal{\textit{et al.}}\fi
\ifx \binstitute  \undefined \def \binstitute#1{#1}\fi
\ifx \bctitle  \undefined \def \bctitle#1{#1}\fi
\ifx \beditor  \undefined \def \beditor#1{#1}\fi
\ifx \bpublisher  \undefined \def \bpublisher#1{#1}\fi
\ifx \bbtitle  \undefined \def \bbtitle#1{\textit{#1}}\fi
\ifx \bedition  \undefined \def \bedition#1{#1}\fi
\ifx \bseriesno  \undefined \def \bseriesno#1{\textbf{#1}}\fi
\ifx \blocation  \undefined \def \blocation#1{#1}\fi
\ifx \bsertitle  \undefined \def \bsertitle#1{\textit{#1}}\fi
\ifx \bsnm \undefined \def \bsnm#1{#1}\fi
\ifx \bsuffix \undefined \def \bsuffix#1{#1}\fi
\ifx \bparticle \undefined \def \bparticle#1{#1}\fi
\ifx \barticle \undefined \def \barticle#1{}\fi
\ifx \botherref \undefined \def \botherref #1{}\fi
\ifx \url \undefined \def \url#1{\textsf{#1}}\fi
\ifx \bchapter \undefined \def \bchapter#1{}\fi
\ifx \bbook \undefined \def \bbook#1{}\fi
\ifx \bcomment \undefined \def \bcomment#1{#1}\fi
\ifx \oauthor \undefined \def \oauthor#1{#1}\fi
\ifx \citeauthoryear \undefined \def \citeauthoryear#1{#1}\fi
\def \endbibitem {}

\bibitem[\protect\citeauthoryear{{Charbonneau}}{2005}]{Dynamo}
\begin{barticle}
\bauthor{\bsnm{{Charbonneau}}, \binits{P.}}:
\byear{2005},
\bjtitle{Living Rev. Solar Phys.}
\bvolume{2},
\bfpage{2}.
\end{barticle}
\endbibitem

\bibitem[\protect\citeauthoryear{{Hathaway}}{2010}]{Models2}
\begin{botherref}
\oauthor{\bsnm{{Hathaway}}, \binits{D.H.}}:
2010,
\it{Does the Current Minimum Validate (or Invalidate) Cycle Prediction
  Methods?},
\textit{ASP Conf. Ser.}
\textbf{428},
307.
\end{botherref}
\endbibitem

\bibitem[\protect\citeauthoryear{{Hathaway}, {Wilson}, and
  {Reichmann}}{1993}]{Models1}
\begin{botherref}
\oauthor{\bsnm{{Hathaway}}, \binits{D.H.}}, \oauthor{\bsnm{{Wilson}},
  \binits{R.M.}}, \oauthor{\bsnm{{Reichmann}}, \binits{E.J.}}:
1993,
\it{The Shape of the Solar Sunspot Cycle},
\textit{Bull. Am. Astron. Soc.}
\textbf{25},
1216.
\end{botherref}
\endbibitem

\bibitem[\protect\citeauthoryear{{Hoyt} and {Schatten}}{1995}]{Hoyt1995}
\begin{barticle}
\bauthor{\bsnm{{Hoyt}}, \binits{D.V.}}, \bauthor{\bsnm{{Schatten}},
  \binits{K.H.}}:
\byear{1995},
\bjtitle{Solar Phys.}
\bvolume{160},
\bfpage{393}.
\end{barticle}
\endbibitem

\bibitem[\protect\citeauthoryear{{Hoyt} and
  {Schatten}}{1998{\natexlab{a}}}]{Hoyt1}
\begin{barticle}
\bauthor{\bsnm{{Hoyt}}, \binits{D.V.}}, \bauthor{\bsnm{{Schatten}},
  \binits{K.H.}}:
\byear{1998},
\bjtitle{Solar Phys.}
\bvolume{181},
\bfpage{491}.
\end{barticle}
\endbibitem

\bibitem[\protect\citeauthoryear{{Hoyt} and
  {Schatten}}{1998{\natexlab{b}}}]{Hoyt2}
\begin{barticle}
\bauthor{\bsnm{{Hoyt}}, \binits{D.V.}}, \bauthor{\bsnm{{Schatten}},
  \binits{K.H.}}:
\byear{1998},
\bjtitle{Solar Phys.}
\bvolume{179},
\bfpage{189}.
\end{barticle}
\endbibitem

\bibitem[\protect\citeauthoryear{{James}}{2007}]{MagneticStorm}
\begin{barticle}
\bauthor{\bsnm{{James}}, \binits{C.R.}}:
\byear{2007},
\bjtitle{Sky Telesc.}
\bvolume{114}(\bissue{1}),
\bfpage{24}.
\end{barticle}
\endbibitem

\bibitem[\protect\citeauthoryear{{Usoskin}}{2008}]{Activity}
\begin{barticle}
\bauthor{\bsnm{{Usoskin}}, \binits{I.G.}}:
\byear{2008},
\bjtitle{Living Rev. Solar Phys.}
\bvolume{5},
\bfpage{3}.
\end{barticle}
\endbibitem

\bibitem[\protect\citeauthoryear{{Waldmeier}}{1961}]{Waldmeier}
\begin{bbook}
\bauthor{\bsnm{{Waldmeier}}, \binits{M.}}:
\byear{1961},
\bbtitle{\it{The Sunspot-Activity in the Years 1610--1960}},
\bpublisher{Schulthess},
\blocation{Zurich}.
\end{bbook}
\endbibitem

\bibitem[\protect\citeauthoryear{{Wilson}}{1995}]{Wilson}
\begin{barticle}
\bauthor{\bsnm{{Wilson}}, \binits{R.M.}}:
\byear{1995},
\bjtitle{Solar Phys.}
\bvolume{158},
\bfpage{197}.
\end{barticle}
\endbibitem

\bibitem[\protect\citeauthoryear{{Wilson} and {Hathaway}}{2005}]{NASA1}
\begin{botherref}
\oauthor{\bsnm{{Wilson}}, \binits{R.M.}}, \oauthor{\bsnm{{Hathaway}},
  \binits{D.H.}}:
2005,
\textit{NASA Technical Report NASA/TP-2005-213608}.
\end{botherref}
\endbibitem

\bibitem[\protect\citeauthoryear{{Wilson} and {Hathaway}}{2006}]{NASA2}
\begin{botherref}
\oauthor{\bsnm{{Wilson}}, \binits{R.M.}}, \oauthor{\bsnm{{Hathaway}},
  \binits{D.H.}}:
2006,
\textit{NASA Technical Report NASA/TP-2006-214601}.
\end{botherref}
\endbibitem

\bibitem[\protect\citeauthoryear{{Wilson} and {Hathaway}}{2007}]{NASA3}
\begin{botherref}
\oauthor{\bsnm{{Wilson}}, \binits{R.M.}}, \oauthor{\bsnm{{Hathaway}},
  \binits{D.H.}}:
2007,
\textit{NASA Technical Report NASA/TP-2007-215134}.
\end{botherref}
\endbibitem

\bibitem[\protect\citeauthoryear{{Wilson} and {Hathaway}}{2009}]{NASA4}
\begin{botherref}
\oauthor{\bsnm{{Wilson}}, \binits{R.M.}}, \oauthor{\bsnm{{Hathaway}},
  \binits{D.H.}}:
2009,
\textit{NASA Technical Report NASA/TP-2009-216061}.
\end{botherref}
\endbibitem

\end{thebibliography}
%
% Without BibTeX 
% \begin{thebibliography}{}
% \bibitem[\protect\citepauthoryear{Author}{Year}]{key}
%   <bibliographical entry>
%
% \bibitem[\protect\citepauthoryear{}{}]{}
%   
%  
% \end{thebibliography}

\end{article} 
\end{document}